\documentstyle[preprint,aps]{revtex}
\tightenlines
\begin{document}

\draft
\title{Testing the Equivalence Principle in Quantum Physics}
\author{Thierry F. Kauffmann}
\address{Department of Physics, Purdue University 1396, West Lafayette
        IN 47907}
\date{}
\maketitle

\begin{abstract}
We showed in a previous paper that a wide class of nonmetric theories of 
gravity encompassed by the $\chi g$ formalism predict that the speed of light 
rays depends on the their polarization direction relative to directions 
singled out by the gravitational field. This gravity-induced birefringence of 
space is a violation of the Equivalence Principle and is due to the nonmetric
coupling between gravity and electromagnetism. In this paper we analyze the
propagation of light in the gravitational field of a rotating black hole 
when nonmetric couplings to curvature are included and compute the time delay 
between rays with orthogonal polarizations. We obtain an upper bound on the 
strength of QED like and CP violating curvature couplings using time delay 
data for pulsar PSR 1937+21. By comparison the corresponding coupling strength
for QED coupling is 37 orders of magnitude less.

\end{abstract}
\section{Introduction}
What distinguishes metric from nonmetric theories is the manner in
which they couple matter to the gravitational field.  Metric
theories admit representations in which a single symmetric, second-rank
tensor gravitational field couples to matter in a specific way that
assures the validity of the Einstein equivalence principle, so-called
universal coupling. In contrast, typical nonmetric theories feature
additional gravitational fields that also couple directly to matter.
For example, Moffat's nonsymmetric gravitation theory (NGT) features an
antisymmetric tensor gravitational field that couples to the electromagnetic
field. Couplings to fields like this violate the Einstein equivalence 
principle by causing local nongravitational physics to depend on the 
additional field values at different events in spacetime.
Ni \cite{Ni77} introduced the $\chi g$ Lagrangian governing
electrodynamics in arbitrary gravitational fields governed by any
of a broad range of metric and nonmetric theories.  The coupling between
gravity and electromagnetism is represented by a fourth rank tensor, $\chi$.
Symmetries of the electromagnetic Lagrangian imply that only twenty-one
components are independent. In this setting, the challenge facing gravitation 
physicists is to design and perform experiments whose results either force this
Lagrangian toward the metric form or reveal the effects of departures
from that form.

We begin by recalling results we obtained for classical electromagnetism.
A first paper \cite{HK1} dealt with a phenomenon that is a reflection
of spatial anisotropy induced by nonmetric gravitational couplings to
the electromagnetic field. This anisotropy can induce a birefringence
of space. Two orthogonal linear polarization states of light are
singled out by a nonmetric field and propagate with different phase
velocities. This effect was first noticed by Gabriel {\it et al}.\
\cite{Gabriel1} in a particular nonmetric theory, Moffat's NGT. He and his
collaborators used observations of polarized light emitted from magnetically
active regions near the Sun's limb to impose sharp new constraints
on that theory. Our first paper interpreted this new type of equivalence 
principle
test in the broader context of the $\chi g$ formalism. It was shown that ten
of the $\chi g$ formalism's twenty-on nonmetric degrees of freedom induce
birefringence, a fact known to Ni. Also, data on the polarization 
of radio galaxies were used to impose constraints that are $10^8$ times sharper 
than previous ones \cite{Krisher}.

In a second paper \cite{HK2} we present the first detailed analysis of the 
constraints which atomic anisotropy experiments impose on all of the 
nonmetric degrees of freedom encompassed by Ni's formalism. It also 
establishes that results of Hughes-Drever-like \cite{HD} experiments, 
in concert with results of birefringence experiments, sharply constrain 
all but two of the nonmetric degrees of freedom encompassed by the $\chi g$ 
formalism. The E\"otv\"os experiment constrains one of the remaining two.

In summary, the combined tests of atomic anisotropy and birefringence of
space impose constraints on all degrees of freedom representing the coupling of gravity with classical electromagnetism, but one. The sharpness of those
constraints strongly supports the validity of the Einstein equivalence
principle.

Things change drastically however when the electromagnetic field is treated as
a quantum field. Photons can exist as virtual electron positrons pairs. Such
pairs have a spatial extension on the order of the Compton wavelength of
electron, and are subject to tidal forces due to background curvature of space.
This effect goes beyond considerations pertaining to the Einstein Equivalence
principle. Local classical theories of matter have no curvature couplings.
This particular coupling between curvature and the electromagnetic
field can lead to effects that mimic violations of the principle of
equivalence because the propagation of photons is directly affected by
the existence of preferred directions in space singled out by the curvature
of the external gravitational field.

Birefringence and other effects induced by the interaction of a quantum
field with a prescribed external field is a phenomenon that has been studied
before. Euler and Heisenberg \cite{EH} computed the effective action
describing the nonlinear interaction of an electromagnetic field with itself
through vacuum polarization. This action was obtained subsequently by
Schwinger \cite{Schwinger} using the background field method he pioneered.
Adler \cite{Adler} used Schwinger's expression to study the quantum
propagation of light in an external magnetic field. He found that the local
speed of light depends on its polarization direction relative to the
orientation of the magnetic field. The effective action governing the
propagation of light in a background gravitational field was derived by
de Wit, using and extending the background field method. This action is
the starting point for studying effects of curvature coupling to
electromagnetism.

The breaking of local Lorentz invariance due to curvature coupling 
causes a background dependence of the local velocity of light, i.e. 
curvature-induced birefringence.
Recently, Drummond and Hathrell \cite{DH} studied the propagation of light
in the gravitational field of a black hole and in cosmological geometries.
For background geometries that are not isotropic, they found photon
trajectories and polarizations for which the local speed of light differs
from $c$. Daniels and Shore \cite{DS1,DS2} performed a similar analysis
for charged and rotating black holes, and found birefringence for
radial photon trajectories in the latter case. This can be understood from the
fact that the gravitational field of rotating black holes singles out the
direction of their axis of rotation.

The coupling of gravity to electromagnetism via curvature is conceptually
very different from the couplings  considered in previous papers. On the
other hand, the effective action describing curvature coupling can be cast
into a form that is encompassed by general Ni's Lagrangian introduced above.

The strength of the curvature coupling between QED and gravity is proportional
to the ratio of the fine structure constant to the mass 
of the electron squared. More general curvature couplings
of the type exhibited by QED in gravitational background can be considered 
for which the parameters setting the scale for the strength of the
coupling are free. One can use observations to restrict their possible values.
Pulses of light of different polarization direction emitted by pulsars
reach the Earth with a time delay due to difference in their propagation
velocity. Experimental limits on measurement of time of arrival provides
a constraint on strength of curvature coupling responsible for
birefringence.

When writing down the effective action for curvature coupling between gravity
and electromagnetism, one restricts oneself to couplings with the
electromagnetic field strength. One could also consider the dual of the
field strength. The reason that possibility is discarded is that such
coupling violates parity and time reversal. One can however conceive the
possibility for quantized gravitational field of coupling to electromagnetism
in a way that is not invariant upon space and time inversions. 
Prasanna and Mohanty \cite{PM} introduce parity and time reversal violating
coupling between photons and gravitons and compute the time delay between 
circularly-polarized light pulses propagating radially from a pulsar to Earth. 
They use data from pulsar PSR 1937+21 to impose upper limit on one parity 
violating coupling coefficient. Their resulting time delay is linear in 
angular momentum of the rotating pulsar, instead of being quadratic as for 
QED like curvature couplings. In addition, they find that the polarization 
modes singled out by the gravitational field are circular polarization, 
whereas the study presented in our paper on birefringence clearly shows 
that these must be linearly polarized.

In this paper we compute the time delay for light rays for QED like and 
CP violating curvature coupling using Ni's formalism and impose a constraint 
on parameter governing the strength of the coupling.
In practice, the strongest constraint on curvature coupling will come
from experiments that search for anisotropy induced by strong gravitational
fields. Such fields exist in the vicinity of black holes or pulsars. Light
pulses of different polarization propagating away from those compact objects 
reach the Earth with a time delay due to their different velocity. This
time delay is a cumulative effect and existing pulsar data are used to 
constrain the strength of parameters entering general curvature coupling.

Because anisotropy in the propagation of light is due to the existence of
preferred directions selected by the gravitational field, birefringence
will be present for rays emanating radially from compact objects whose 
gravity field is not spherically symmetric. Rotating objects such as pulsars
single out the direction of their axis of rotation and are therefore good
candidates for birefringence effects that interest us. 
We obtain the relative velocity difference between light pulses of different
polarization for QED like and CP violating coupling. We correct a mistake 
in Daniels and Shore's computation, which does not affect their result in 
a significant way. Our expression in the CP violating case involves the 
square of the angular momentum, as expected.
We then integrate the relative difference in speed of light along a ray
emanating from the surface of the pulsar and reaching the Earth. Comparison
with data from PSR 1937+21 yields an upper bound on the strength of general
curvature coupling.

\section{Photon propagation in Kerr geometry}

The equations of motion for the electromagnetic field when subject to general
curvature coupling with gravity are obtained from

\begin{equation}
 \frac{\delta \Gamma}{\delta A_\mu(x)} = 0
 \label{}
\end{equation}
where the effective action $\Gamma$ decomposes into
\begin{equation}
 \Gamma_0 = -\frac{1}{16\pi}\int d^4\, x\sqrt{-g} F_{\mu\nu}F^{\mu\nu}
\end{equation}
and $\Gamma_1$ represents general curvature coupling of the type 
exhibited by one-loop QED in a gravitational field 
\begin{equation}
 \Gamma_1 = -\frac{1}{16\pi}\int d^4\, x\sqrt{-g} 
 \big( \alpha RF_{\mu\nu}F^{\mu\nu} + \beta R_{\mu\nu}F^{\mu\lambda}
  F^\nu{}_\lambda + \sigma R_{\mu\nu\lambda\rho} F^{\mu\nu}F^{\lambda\rho} \big)
\end{equation}
The corresponding Lagrangian is of the form
\begin{equation}
 {\cal L}_{eff} = -\frac{1}{16\pi}\chi^{\alpha\beta\gamma\delta}
                 F_{\alpha\beta} F_{\gamma\delta}
\end{equation}
Where $\chi^{\alpha\beta\gamma\delta}$ is expressed in terms of the Riemann
tensor and its contractions. We evaluate it for the gravitational field
created by a rotating compact object.

The geometry outside a rotating compact object such as a pulsar or black hole, 
is described in general relativity by the Kerr metric. Even though curvature 
couplings are outside general relativity, they represent a first order 
correction and are assumed not to lead to significant modifications to the
Kerr solution. The metric outside a pulsar depends on its mass $M$ and angular 
momentum $Ma$, expressed in geometrized units where $c=G=1$. It describes the 
classical final state of the collapse of a rotating electrically neutral 
object. 
One important feature of the Kerr metric is that it contains a non diagonal
component $g_{\phi t}$ expressing the dragging of inertial frames at the 
horizon at the angular velocity 
\begin{equation}
 \omega = \frac{2aMr}{\Sigma^2} 
\end{equation}
where

\begin{equation}
 \Sigma^2 = (r^2 + a^2)^2 - a^2 \Delta \sin^2\theta\qquad 
 \Delta = r^2 - 2Mr + a^2
\end{equation}
It will be necessary to compensate for this rotation when introducing 
a local Lorentz coordinate system.

In analysis of birefringence of light presented in chapter two, we expressed 
the relative difference in light velocity between different polarizations 
in function of values of coupling tensor in local quasi-Lorentz frame

\begin{equation}
	\frac{\delta c}{c} = \frac{1}{2}\sqrt{(A-C)^2 + 4B^2}
\end{equation}
where we recall
\begin{eqnarray}
 A &=& \delta\chi^{0303} - 2\delta\chi^{0331} - \delta\chi^{3131} \nonumber\\
 B &=& \delta\chi^{0203} + \delta\chi^{0312} + \delta\chi^{0213} + 
       \delta\chi^{3112} \nonumber\\
 C &=& \delta\chi^{0202} - 2\delta\chi^{0221} - \delta\chi^{2121}
\end{eqnarray}
in a frame where the direction of light propagation is $\hat r$, and
\begin{equation}
 \chi^{\alpha\beta\gamma\delta} = \chi_{Lorentz}^{\alpha\beta\gamma\delta}
 + \delta\chi^{\alpha\beta\gamma\delta}
\end{equation}

Since outside the black hole the curvature scalar and the Ricci tensor vanish,
one has
\begin{equation}
 \delta\chi^{\alpha\beta\gamma\delta} = \sqrt{-g}\sigma 
  R^{\alpha\beta\gamma\delta}
\end{equation}

The coordinate transformation between Kerr and local Lorentz coordinates is
\begin{eqnarray}
 t' &=& t\sqrt{g_{00}-\omega^2 g_{\phi\phi}}  \nonumber\\
 \theta' &=& \theta\sqrt{g_{\theta\theta}}  \nonumber\\
 \phi' &=& \sqrt{g_{\phi\phi}} (\phi + \omega t) \nonumber\\
 r' &=& r\sqrt{g_{rr}}
\end{eqnarray}
The Riemann tensor components we need for radial propagation are
\cite{Chandra} \cite{DS2}
\begin{eqnarray}
 R_{0101} = a =\frac{Mr}{\rho^6\Sigma^2}(r^2-3a^2\cos^2\theta)(2(r^2+a^2)^2
 + a^2\Delta\sin^2\theta) \nonumber\\
 R_{0202} = b=-\frac{Mr}{\rho^6\Sigma^2}(r^2-3a^2\cos^2\theta)((r^2+a^2)^2
 + 2a^2\Delta\sin^2\theta) \nonumber\\
 R_{0123} = c=-\frac{Ma\cos\theta}{\rho^6\Sigma^2}(3r^2-a^2\cos^2\theta)
 (2(r^2+a^2)^2+ a^2\Delta\sin^2\theta) \nonumber\\
 R_{0231} = d=\frac{Ma\cos\theta}{\rho^6\Sigma^2}(3r^2-a^2\cos^2\theta)
 ((r^2+a^2)^2+ 2a^2\Delta\sin^2\theta) \nonumber\\
 R_{2323}=-R_{0101}\quad R_{1313}=-R_{0202}\quad R_{1212}=-R_{0303}=R_{0101}
 + R_{0202} \nonumber\\
 R_{0312}=-R_{0123}-R_{0231}\quad R_{3132}=-R_{0102}\quad R_{0223}=-R_{0113}
\end{eqnarray}

Note that, in this local Lorentz frame, indices are raised and lowered using
the Minkowski metric.

With the corresponding expression for $\delta\chi$ in the local tetrad, one 
finds
\begin{equation}
 \frac{\delta c}{c} = \sigma\Delta\frac{3Ma^2\sin^2\theta}{\rho^6\Sigma^2}
 \sqrt{r^2(r^2-3a^2\cos^2\theta)^2 + 
 a^2\cos^2\theta(3r^2-a^2\cos^2\theta)^2}
\end{equation}

This expression differs from that obtained by Daniels and Shore, due to their 
incorrect expression for the Riemann tensor (their Eq. 3.8). 
The velocity shift vanishes on the horizon for each polarization.

\section{Birefringence of light in CP violating interactions with gravity}

When writing down the effective action for quantum electrodynamics in curved
background, one restricts oneself to couplings with the electromagnetic field
strength. On pure dimensional grounds, one could also consider the dual of the
field strength. The reason that possibility is discarded is that such coupling
violate parity. On the other hand, the very existence of black hole
radiance suggests that gravity might violate T invariance \cite{Wald}.
Lack of time reversal invariance occurs because the evolution
from an initial pure state such as the $in$ vacuum toward a mixed state 
characterized by the temperature of the black hole radiation, is irreversible.
Prasanna and Mohanty \cite{PM} are considering possible CP 
violating couplings between photons and gravitons. Dimensional analysis in 
units of mass restricts the form the interaction Lagrangian as follows. 
Keeping terms quadratic in electromagnetic field strength constraints 
interaction 
terms to contain only gravitational expression of mass dimension two. Such 
objects involve at most two derivatives of the metric.
General covariance then forces those to be formed out of local geometrical
invariants such as the Riemann tensor and its contractions. The
resulting expression is then similar to QED like coupling to gravitational
field.

Constructing a Lagrangian describing CP violating graviton-photon coupling
is similar to building {\it ab initio} an effective Lagrangian for one-loop QED
in curved spacetime. One writes all gauge invariant and diffeomorphism
invariants operators based on field strength and curvature tensor. Those 
operators can be ordered by increasing mass dimension. In the case of CP
violating interaction, the lowest dimensional operator has dimension four 
and is $F^{\mu\nu}{\tilde F} _{\mu\nu}$.  Since it is a total divergence 
in four dimensional spacetime, it plays no dynamical role in the absence 
of boundaries. The next terms all have dimension six, thus
\begin{equation}
 {\cal L}_{int} = -\frac{1}{16\pi}\sqrt{-g}
  (F_{\mu\nu}F^{\mu\nu}+c_1 R_{\mu\nu\alpha\beta}F^{\mu\nu}
 {\tilde F}^{\alpha\beta} + c_2 R_{\mu\nu}F^{\mu\alpha}{\tilde F}^\nu{}
 _\alpha + c_3 R F^{\mu\nu}{\tilde F}_{\mu\nu})
\end{equation}
where ${\tilde F}_{\mu\nu} = 1/2 \epsilon^{\alpha\beta\mu\nu}F_{\alpha\beta}$
is the dual of $F$.All coefficients $c_i$ have thus mass dimension minus two.
The last six dimensional interaction term involves one derivative of $F$ and 
$\tilde F$ and does not couple to gravity.

Outside the pulsar, both the Ricci tensor and the curvature scalar vanish, and
the resulting interaction Lagrangian only involves the indeterminate 
coefficient $c_1$. From the expression for the CP violating Lagrangian one 
extracts the gravitational tensor $\delta\chi$ 
\begin{equation}
 \delta\chi^{acef} = \frac{c_1}{2}R^{acbd}\epsilon_{bd}{}{}^{ef} 
\end{equation}
For this Lagrangian, the relative difference of light velocity is
\begin{eqnarray}
 \frac{\delta c}{c} = c_1\Delta\frac{3Ma^2\sin^2\theta}{\rho^6\Sigma^2}
 \sqrt{r^2(r^2-3a^2\cos^2\theta)^2 +
 a^2\cos^2\theta(3r^2-a^2\cos^2\theta)^2}
\end{eqnarray}

The phase shift is thus identical to the one obtained in QED.
 Note that this expression does vanish on the horizon, just like in the QED 
 case.Therefore the CP violating coupling satisfies the horizon theorem 
 \cite{Shore} which says that at the event horizon, (where 
 $\Delta$ vanishes), the light cone is not modified by the coupling with the
 background geometry.This theorem holds at the classical level and for one 
 loop QED in curved spacetime.
 
\section{Experimental constraints}
We use time measurements from pulsars to constrain the magnitude of curvature
coupling coefficient. For such rotating objects, lowest order contribution in
angular momentum parameter $a$ to the velocity shift gives reasonable order 
of magnitude for comparisons with observations. One finds
\begin{equation}
 \frac{\delta c}{c} = c_1\frac{3Ma^2\sin^2\theta}{r^5}(1-\frac{2M}{r}) 
 + O(\frac{c_1 M a^4}{r^7})
\end{equation}

To obtain the integrated phase shift, we must express the relative velocity
shift as a coordinate quantity. That is the expression we just derived measures
ratios of local Lorentz length and time intervals. The measured time delay
at infinity differs from that measured in the vicinity of the black hole.
The actual phase shift must thus be multiplied by 
$\sqrt{\frac{g_{rr}}{-g_{00}}}$. Since the local velocity shift is a second 
order quantity in $a$, we only keep the $a$ independent terms in the metric 
coefficients, i.e we use the Schwarzschild metric. The integrated time delay 
from the surface of the pulsar to infinity yields 
\begin{equation}
 \Delta t = \int_{R}^\infty c_1\frac{3Ma^2\sin^2(\theta)}{r^5}
 (1-\frac{2M}{r})^2 \, dr
\end{equation}
To obtain an order of magnitude for the strength of CP violating curvature
coupling one can choose an equatorial ray, for which one obtains.
\begin{equation}
 \Delta t = \frac{1}{20}c_1 \frac{J^2 G}{c^5 M R^4}
\end{equation}
The time delay for pulsar PSR 1937+21 is less than $10^{-6}s$. 
For this pulsar, of angular momentum $J=Mac=10^{41}kg m^2 s^{-1}$,
radius $R=10 km$
and mass $M= 1.4 M_{\odot}$ one finds
the following constraint for $c_1$
\begin{equation}
 c_1 \le 10^{9} m^2 
\end{equation}
The same constraint on QED like curvature coupling is obtained, since 
expression for the relative velocity shift is identical for rays propagating
radially.

\section{Summary and conclusions}

We have derived the time delay of light emitted from pulsars as it travels
in a gravitational field when that field couples to electromagnetism through its
Riemann tensor. An upper bound on the strength of both QED like and CP 
violating curvature coupling is obtained by computing the time delay along
equatorial rays and using data from pulsar PSR 1937+21. 
By comparison the corresponding value of $\sigma$ for QED in curved spacetime
is 37 orders of magnitude less.

 \end{document}